\documentclass[12pt]{article}

\usepackage{amsmath,amssymb,amsfonts}
\usepackage[paper=letterpaper,margin=0.9in]{geometry}

\newcommand{\be}{\begin{equation}}
\newcommand{\ee}{\end{equation}}
\newcommand{\bea}{\begin{eqnarray}}
\newcommand{\eea}{\end{eqnarray}}
\newcommand{\ba}{\begin{array}}
\newcommand{\ea}{\end{array}}

\newcommand{\tr}[1]{{\rm tr}\, {1}}
\newcommand{\bra}[1]{\langle{1}|}
\newcommand{\ket}[1]{|{1}\rangle}
\newcommand{\ip}[2]{\langle{1}|{2}\rangle}
\newcommand{\vev}[1]{\langle{1}\rangle}

\newcommand{\todo}[1]{{\bf {1}}}
\newcommand{\comment}[1]{}

\newcommand{\ft}[2]{{\textstyle\frac{1}{2}}}
\def\fract12{{\textstyle{1\over2}}}
\def\ffract12{\raise .3 em\hbox{$\scriptstyle1$}\kern-.25em/
                \kern-.2em\lower .2 em \hbox{$\scriptstyle2$}}
\def\fractje12{{\scriptstyle{1\over2}}}

\def\part12{{\partial1\over\partial2}}
\def\ex1{e^{\textstyle1}}

\begin{document}

\renewcommand{\thepage}{\arabic{page}}
\setcounter{page}{1}


\vskip 2.00 cm
\renewcommand{\thefootnote}{\fnsymbol{footnote}}

\centerline{\bf \Large A Precision Test for an Extra Spatial Dimension}
\centerline{\bf \Large Using Black Hole--Pulsar Binaries}

\vskip 0.75 cm

\centerline{{\bf
John H. Simonetti${}^{1}$\footnote{\tt jhs@vt.edu}
Michael Kavic,${}^{2}$\footnote{\tt michael.kavic@liu.edu}
Djordje Minic${}^{1}$\footnote{\tt dminic@vt.edu} 
Umair Surani${}^{1}$\footnote{\tt usurani@vt.edu } and
Vipin Vijayan${}^{1}$\footnote{\tt vvijayan@nd.edu }
}}
\vskip .5cm
\centerline{${}^1$\it Department of Physics}
\centerline{\it Virginia Tech}
\centerline{\it Blacksburg, VA 24061, U.S.A.}
\vskip .5cm
\centerline{${}^2$\it Department of Physics}
\centerline{\it Long Island University}
\centerline{\it Brooklyn, NY 11201, U.S.A.}
\vskip .5cm

\setcounter{footnote}{0}
\renewcommand{\thefootnote}{\arabic{footnote}}

\begin{abstract}
We discuss the observable effects of enhanced black-hole mass loss in a black hole--neutron star (BH--NS) binary, due to the presence of a warped extra spatial dimension of curvature radius $L$ in the braneworld scenario.  For some masses and orbital parameters in the expected ranges the binary components would outspiral, the opposite of the behavior due to energy loss from gravitational radiation alone.  If the NS is a pulsar, observations of the rate of change of the orbital period with a precision obtained for the Binary Pulsar B1913+16 could easily detect the effect of mass loss.   For $M_{BH}=7M_\odot$, $M_{NS}=1.4M_\odot$, eccentricity $e=0.1$, and $L=10\mu$m, the critical orbital period dividing systems which inspiral from systems which outspiral is P$\approx$6.5~hours, which is within the range of expected orbital periods; this value drops to P$\approx$4.2~hours for $M_{BH}=5M_\odot$.  Observations of a BH--pulsar system could set considerably better limits on $L$  in these braneworld models than could be determined by torsion-balance gravity experiments in the foreseeable future.
\end{abstract}


\newpage

\section{Introduction}

Extra spatial dimensions, beyond the three encountered in everyday experience, have long been discussed in theoretical physics \cite{Kaluza, Klein, Rattazzi2006}.  Earth-based tests of these ideas are difficult \cite{Adelberger:2009zz}.   We discuss an astrophysical test for the consequences of a particular class of extra spatial dimension models.  The observations and analysis would be similar to work done on the Binary Pulsar PSR~B1913+16 (two neutron stars, one observed as a pulsar), which provided a high-precision test of the effect of energy loss by gravitational radiation \cite{Taylor:1982zz, Taylor:1989sw, Weisberg:2004hi}.  In the test presented here the binary pair would consist of a black hole (BH) and a neutron star (NS) observed as a pulsar.  The intended audience for this paper comprises both astronomers familiar with observational astrophysics but unfamiliar with extra dimensions, and theoretical physicists in the opposite predicament.  We hope that this paper and the similar works which have proceeded it \cite{Emparan:2002px, PsaltisXTEJ1118, Johannsen2009, JohannsenPsaltisMcClintock2009, McWilliams:2009ym} will help these two groups to bridge the gap between them, and demonstrate that there are potential astrophysical avenues for progress in this area.

One reason for interest in extra dimensions is string theory, the most notable quantum gravity theory, which requires 6 or 7 extra spatial dimensions \cite{polch}. Such extra dimensions could evade detection if they are extremely small --- ``compactified'' or rolled-up; a coordinate $r$ along such a dimension is periodic on a length scale $L$, i.e., $r = r +L$.  The most natural scale for these dimensions is the Planck length $L\sim\sqrt{\hbar G/c^3}\sim10^{-33}$~cm.  Probing that length scale in accelerators would require energies at the Planck energy $\sim\sqrt{\hbar c^5/G}\sim10^{19}$~GeV,  $10^{16}$ times larger than the electroweak TeV scale of current experiments \cite{LHC}.

A second motivation for considering extra spatial dimensions is the ``hierarchy problem'' --- the relative weakness of gravity compared to the other three fundamental forces.  If gravitons alone propagate in extra spatial dimensions, then the gravitational field of a particle drops faster than $1/r^2$ on length scales smaller than the size of an extra dimension, and is therefore weaker on larger scales where the inverse-square-law behavior becomes manifest.  In addition, $L$ could be $\sim1$~mm and be undetected in particle accelerator experiments (which don't probe gravity), and torsion-balance gravity experiments \cite{ADD, I1}.  Gravitational torsion balance experiments have set a limit of $L< 44\mu$m (95\% confidence) \cite{Kapner2007}.  \cite{Adelberger:2009zz} state that modest improvements in these torsion balance experiments will occur, but their own projections show sensitivity far beyond the current limits may prove difficult in such experiments.

An alternative to compactification is possible. Extra dimensions of infinite size are allowable, if their geometry is ``warped'' so gravity cannot propagate further than the length scale set by the torsion-balance experiments.  For one class of models all forces and particles other than gravitons exist only on a ``brane,'' which is the boundary of a infinite, warped ``bulk''  (e.g., the Randall-Sundrum 2 model, RS2) \cite{RS2}.  In this ``braneworld'' scenario, our 3$+$1 dimensional world is the brane, populated with the standard-model particles and forces.  An extra spatial dimension is ``orthogonal'' to the brane, forming the bulk. In such warped models the bulk must be an anti de Sitter (AdS) space. Gravitons propagate on the brane and in the bulk, thus, in principle, resolving the hierarchy problem.  The ``warping'' amounts to a redshifting of gravity as it propagates into the bulk, so, in effect, gravitational experiments only probe a short distance into the bulk.

Much work has been done to understand the nature of black holes in the braneworld scenario \cite{Gregory:2008rf}.  Because the bulk space is anti de Sitter space, application of the AdS/CFT conjecture \cite{Maldacena:1997re} becomes possible, particularly in the context of RS2.  The result is that the full classical 5D braneworld black hole solution is equivalent to a 4D quantum corrected black hole.  Moreover, this analysis yields a dramatically increased evaporation rate for large black holes due to the existence of conformal degrees of freedom \cite{Emparan:2002jp}.   This conclusion was challenged by \cite{Fitzpatrick2006} in which a static black hole solution was demonstrated to exist. This solution however was found to be unstable. Moreover, numerical investigations have failed to yield evidence of a stable static macroscopic black hole in the context of the braneworld scenario \cite{Kudoh:2003xz, Yoshino:2008rx}. The lack of a dynamically stable solution may serve to indicate that the RS2 model implies a dramatic increase in the amount of Hawking radiation emitted by macroscopic black holes.

Note that we can give a broader heuristic argument for the general significance of quantum gravity in relativistic astrophysical situations involving black holes, without tying ourselves directly to specific models involving extra dimensions. On general grounds \cite{gressay, Kavic:2008qb}, quantum gravity, due to its essential non-locality dictated by the underlying classical symmetry of general coordinate transformations, can be argued to be sensitive to other scales, such as the TeV scale of particle physics, where, also, by unitarity, new physical phenomena are expected to appear. Similarly, the increased evaporation rate can be expected based on models of black holes as bound states of these essentially non-local quantum gravitational degrees of freedom. Such models often arise in the analysis of black hole evaporation in string theory \cite{mathur}, in which the non-local, stringy, quantum gravitational degrees of freedom can spread from the Planck scale to large, astrophysically significant scales. Thus a very large number of degrees of freedom can be transferred from very small to very large length scales, providing an enhanced evaporation rate that is crucial for the discussion that follows.

\section{Astronomical Limits}

Some observable astrophysical implications of these results were explored by \cite{Emparan:2002px}.  Observations of X-ray binary systems (BH $+$ companion star) can potentially constrain the size of extra dimensions in the braneworld scenario.  
The constraints rely on the kinematic age for the system or observational limits on any change in the orbital period of the system. These observations have the potential to set lower limits than can be obtained by current torsion-balance experiments.  Using the kinematic age of the X-ray binary XTE J1118$+$480 \cite{PsaltisXTEJ1118} set an upper limit of $L< 80\mu$m (95\% confidence).  \cite{Johannsen2009} set an upper limit of $L< 970\mu$m (3$\sigma$) from current observational limits on the change in orbital period of XTE J1118$+$480, while \cite{JohannsenPsaltisMcClintock2009} set a limit of $L< 161\mu$m (3$\sigma$) from observations of A0620$-$00.  \cite{Johannsen2009} states that one additional measurement of the orbital period of XTE J1118$+$480 could constrain $L$ to less than 35$\mu$m.  However, the physical complexity of such systems and our limited understanding of their astrophysical behavior makes a proper interpretation of the observations difficult, and perhaps ambiguous.  While torsion-balance experiments may not be able to set such low limits, they provide cleaner experimental setups, with a more straight-forward interpretation.

Cleaner astrophysical binary systems comprise components that can be treated as point masses, with no mass exchange. \cite{McWilliams:2009ym} discusses a BH--BH or BH--NS system with enhanced BH evaporation, as it would be observed by a gravity wave detector such as the proposed Laser Interferometer Space Antenna (LISA).  As we will discuss below, BH mass loss can lead to an outspiral of the binary components, while the effect of gravitational radiation produces inspiral.  As discussed by McWilliams, for expected binary system parameters a changing orbital period (i.e., chirping of the gravity wave signal) could not be measured by LISA, thus one could not directly confirm an outspiral behavior.  However, if one assumes any observed binary of sufficiently small orbital period has reached that period by inspiraling, then one can set a limit on $L$.  For expected periods and masses, McWilliams shows that a limit of $L<5\mu$m can be set.

\section{A Black Hole-Pulsar Binary System in the Presence of a Warped Extra Spatial Dimension}

We consider a binary system consisting of a BH and an NS, where the NS is a pulsar.  Observations of the pulsar could be used to measure the changing orbital period of the system with sufficiently high precision to either directly observe the outspiral behavior, or measure the competing contributions of mass loss and gravitational radiation as they affect the inspiral rate of the system.  Alternatively, improved limits could be set on the size of an extra dimension.  We are motivated by the PSR~B1913+16 Binary Pulsar case, where observations of the one NS that acts as a pulsar have yielded \textit{high precision} determinations of the  parameters of the system, and have provided a dramatic test of relativistic physics.  

The BH evaporation rate in the braneworld scenario is given by
\begin{equation}
\label{massloss}
\dot{M}_{BH}=-2.8\times 10^{-7}M_{BH}^{-2}L_{10}^2 \ M_{\odot}~\textrm{y}^{-1} \, ,
\end{equation}
where $M_{BH}$ is the black hole mass in solar masses, and $L_{10}$ is the AdS radius in units of 10~$\mu$m \cite{Emparan:2002jp, Emparan:2002px}.  
Hereafter, $M$ and $m$ represent masses in units of solar masses and SI units, respectively. The current bound on the AdS radius is $L<44\mu$m \cite{Kapner2007, Adelberger:2009zz} from torsion-balance experiments.  We take  $L=10\mu$m (i.e., $L_{10}=1$) as a nominal value in the discussions that follow.

\subsection{The Effects of Mass Loss Due to Enhanced BH Evaporation}

Consider a BH--NS binary system, with mass loss alone.  We will use the results of Hadjidemetriou \cite{Hadjidemetriou1963, Hadjidemetriou1966}, who analyzed the dynamical behavior of a binary system with isotropic mass loss from one or both components, in the classical case (relativistic corrections will be considerably smaller in magnitude than the results obtained here).  Intuitively, the binary pair becomes less tightly bound, so the components must separate over time and the orbital period will increase.   In a fixed, non-rotating frame of reference, let $m_1$, and $m_2$ be the masses of the two components, and $m = m_1 + m_2$.  The rates of change of the osculating elements for the orbit of $m_2$ relative to $m_1$ are given by
\begin{eqnarray}
\dot{a} &=& -a \frac{1 + 2e \cos \theta + e^2}{1 - e^2} \frac{\dot{m}}{m} = - a \frac{\dot{m}}{m}, \\
\dot{e} &=& - \left( e + \cos \theta \right) \frac{\dot{m}}{m} = 0, \\
\dot{\omega} &=& - \frac{\sin \theta}{e} \frac{\dot{m}}{m} = 0
\end{eqnarray}
where $a$ is the semi-major axis, $e$ is the eccentricity, $\omega$ is the longitude of periastron, and $\theta$ is the true anomaly \cite{Hadjidemetriou1963, Hadjidemetriou1966}.  The results given above are averages over one orbit, assuming $\dot{m}P \ll m$ (as is our case); one orbit is well approximated by a Keplerian orbit where the average of $\cos\theta=-e$ \cite{Sivardiere1985} and the average of $\sin \theta = 0$.

Kepler's third law relates $P$, $a$, and $m$.  Thus, to lowest order the rate of change of the orbital period due to mass loss (ML), averaged over one orbit, is
\begin{equation}
\dot{P}_{ML} = \frac{3}{2} P \frac{\dot{a}}{a} \left( 1 - \frac{1}{3} \frac{\dot{m}}{m} \frac{a}{\dot{a}} \right) = 2 P \frac{\dot{a}}{a}.
\end{equation}

Summarizing the results in terms of the parameters $m$, $\dot{m}$, and $P$, we have $
\dot{e}_{ML} = \dot{\omega}_{ML} = 0$ and
\begin{equation}
\dot{a}_{ML} = - \left( \frac{G}{4\pi^2} \right)^{1/3} \left( \frac{P}{m} \right)^{2/3} \dot{m},
\end{equation}
\begin{equation}
\dot{P}_{ML} = -2 P \frac{\dot{m}}{m}.
\end{equation}
For mass loss, $a$ and $P$ increase.
In the braneworld scenario the mass loss rate for the binary is the mass loss rate from the black hole, given by eq.~(\ref{massloss}). 

\subsection{The Effects of Energy Loss Due to Gravitational Radiation}

The effects of energy loss by gravitational radiation from a binary system are well known \cite{Peters:1963ux, Peters:1964zz}: $a$ and $P$ decrease --- opposite results from the mass-loss scenario.  For two point masses (i.e., $a\gg$ radius of either object), and where averages over one orbit are of sufficient accuracy, from \cite{Peters:1964zz}, and \cite{Taylor:1989sw} (in the context of the PSR~B1913+16),  we have
\begin{equation}
\dot{a}_{GR} = - \frac{256\pi^2}{5} \frac{G^2}{c^5} \frac{m_1 m_2}{P^2}
\frac{\left( 1 + \frac{73}{24}e^2 + \frac{37}{96}e^4 \right)}{\left( 1-e^2 \right)^{7/2}},
\end{equation}
\begin{equation}
\dot{P}_{GR} = - \frac{192\pi}{5} \frac{\left( 2\pi G \right)^{5/3}}{c^5} \frac{m_1 m_2}{m^{1/3}P^{5/3} } 
\frac{\left( 1 + \frac{73}{24}e^2 + \frac{37}{96}e^4  \right)}{\left( 1-e^2 \right)^{7/2}} , 
\end{equation}
\begin{equation}
\dot{e}_{GR} = - \frac{304}{15} \left( 4\pi^2 \right)^{4/3}  \frac{G^{5/3}}{c^5} \frac{m_1 m_2}{M^{1/3} P^{8/3}}
\frac{e \left( 1 + \frac{121}{304}e^2  \right)}{\left( 1-e^2 \right)^{5/2} }, 
\end{equation}
\begin{equation}
\dot{\omega}_{GR} = 6\pi \left( 4\pi^2 \right)^{1/3} \frac{G^{2/3}}{c^2} \frac{m^{2/3}}{P^{5/3}} \left( 1-e^2 \right)^{-1}.
\end{equation}

Henceforth, we will concentrate on $\dot{a}$, which is most readily observed as a rate of change of the period $\dot{P}$.

\subsection{The Results for Typical Parameters: Inspiral versus Outspiral}

It is instructive to examine the above results in terms of specific values for the parameters.  Consider a binary comprising a black hole of mass $M_{BH}=3$ (units of $M_\odot$) and neutron star of mass $M_{NS}=1.4$, with orbital period $P_{10}=1$ in units of 10~hours.  The semi-major axis is
\begin{equation}
a = 2.68\times10^9\ {\rm m}\ \left(M_{BH}+M_{NS}\right)
P_{10}
\end{equation}
which is $\sim R_\odot \gg$ radius of either object.

Considering the effect of mass loss only, we find
\begin{eqnarray}
\dot{a}_{ML} &=& 19\ {\rm m\ y}^{-1}
\left( \frac{M_{BH}+M_{NS}}{4.4} \right)^{-2/3} 
\left( \frac{M_{BH}}{3} \right)^{-2} P_{10}^{2/3} L_{10}^2, \\
\dot{P}_{ML} &=& 0.51\ {\rm ms\ y}^{-1} 
\left( \frac{M_{BH}+M_{NS}}{4.4} \right)^{-1}
\left( \frac{M_{BH}}{3} \right)^{-2} P_{10} L_{10}^2,
\label{periodrate}
\end{eqnarray}
which are independent of $e$.

Considering only the effect of gravitational radiation, with no mass loss, we have
\begin{eqnarray}
\dot{a}_{GR} &=& -0.38\ {\rm m\ y}^{-1}
\left( \frac{M_{BH}}{3} \right)
\left( \frac{M_{NS}}{1.4} \right)
P_{10}^{-2}
f(e),\\
\dot{P}_{GR} &=& -0.0076\ {\rm ms\ y}^{-1}
\left( \frac{M_{BH}}{3} \right)
\left( \frac{M_{NS}}{1.4} \right)
\left( \frac{M_{BH}+M_{NS}}{4.4} \right)^{-1/3}
P_{10}^{-5/3} f(e),
\label{GRperiodrate}
\end{eqnarray}
where
\begin{equation}
f(e) = \left( 1 + \frac{73}{24}e^2 + \frac{37}{96}e^4 \right) \left( 1-e^2 \right)^{-7/2}.
\end{equation}
Clearly, for systems with these parameters the enhanced mass loss dominates the evolution of $P$. 

Consider a specific case with $P=7.75$~hours and $e=0.6$, motivated by PSR~B1913+16.  The observed rate of change of the orbital period for the binary pulsar PSR~B1913$+$16 is $\dot{P} = (-2.4184\pm 0.0009)\times10^{-12}$~s/s $= (-0.076\pm0.00003)$~ms~y$^{-1}$  \cite{Weisberg:2004hi}.  For $L_{10}=1$, we find $\dot{a}_{ML} = 16\ {\rm m\ y}^{-1}$, and $\dot{P}_{ML} = 0.40\ {\rm ms\ y}^{-1}$, while the effects due to gravitational radiation are of opposite sign, but roughly 4 times smaller, at $\dot{a}_{GR} = -3.9\ {\rm m\ y}^{-1}$, and $\dot{P}_{GR} = -0.12\ {\rm ms\ y}^{-1}$.  $\dot{P}_{ML} = 0.40\ {\rm ms\ y}^{-1}$ is about 13,000 times larger than the precision on the measured rate of change of the orbital period of PSR~B1913$+$16.  The effect of mass loss could be well-determined by observations of a similar precision.

In a BH--NS system gravitational radiation and BH mass-loss produce opposite effects.  For small $P$, gravitational radiation dominates and $P$ will decrease as the binary components inspiral and eventually coalesce.  For large $P$, BH mass-loss dominates, thus $P$ increases and the components outspiral.   Equating the magnitudes of the rate of change of the period for mass loss only (eq.~\ref{periodrate}) and for gravitational radiation only (eq.~\ref{GRperiodrate}), we obtain an approximate critical orbital period $P_{\rm critical}$ which separates these two disparate behaviors,
\begin{equation}
P_{\rm critical} \approx 2.1\ {\rm hours} 
\left( \frac{M_{BH}}{3} \right)^{9/8}
\left( \frac{M_{NS}}{1.4} \right)^{3/8}
\left( \frac{M_{BH}+M_{NS}}{4.4} \right)^{1/4}
L_{10}^{-3/4}
f(e)^{3/8}.
\end{equation}

\section{Discussion}

If a significant number of binaries have  $P > P_{\rm critical}$ there are implications for discussions of the number of BH--NS  binaries in the Galaxy, since such discussions assume \textit{all} binary systems inspiral and eventually cease to exist.  There would also be implications for gravity wave searches, which depend on the number of inspiraling compact object binaries.

\cite{Pfahl2005} investigated the theoretical formation rate and evolution of BH--NS binaries where the NS is a ``recycled'' pulsar (an increased spin rate from accretion, thus a greater pulsar ``clock'' stability, yielding higher precision observations, e.g., PSR~B1913+16).  They find that most newly formed BH--NS pairs of this sort have an orbital period of 2--7~hours, with a peak near 3~hours, and an eccentricity $e<0.3$, with a peak near 0.1.  The BH masses range from 5--10~$M_\odot$, with an average of $\approx7$~$M_\odot$.

Taking $M_{BH} = 7$, $M_{NS} = 1.4$, $P=3$~hours, $e=0.1$, and $L=10\mu$m, we obtain $\dot{P}_{ML}=0.015$~ms~y$^{-1}$, and $\dot{P}_{GR}=-0.11$~ms~y$^{-1}$.  The effect of gravitational radiation is ten times larger than the effect of mass loss, but precision measurements of the orbital period would show a deviation from the results due to gravitational radiation alone: $\dot{P}_{ML}=0.015$~ms~y$^{-1}$ is well above the obtainable observational precision of $\pm0.00003$~ms~y$^{-1}$ (the precision stated here is specific to the binary pulsar PSR~B1913$+$16 and will vary from case to case).  Note that the critical period for inspiral versus outspiral is 6.5~hours for these masses, eccentricity, and AdS radius, which is within the range of orbital periods expected; it is as low as 4.2~hours for $M_{BH} = 5$.

\cite{Pfahl2005} also determined the expected number of BH--NS binaries with a recycled pulsar in our Galaxy to be less than $\sim$10.   This number is quite small in part because of the expected coalescence rate of systems with periods of 3--4 hours and inspiral lifetimes of $\sim10^8$ years.  But the effect of mass loss, if present, is not accounted for, of course.  BH-NS systems that don't contain a recycled pulsar are also possible, and \cite{Belczynski2002} estimate there are some hundreds of such systems in total.  Nevertheless, the search for BH--NS systems (with an observable pulsar) will be difficult, of course.  However, searches for pulsars in binary systems are of particular interest, and these searches will become more sensitive with the appearance of new instruments, (e.g., the Square Kilometer Array).

Given the prospects for improvement of torsion-balance measurements over the coming decades, it is reasonable to ascertain the prospects for BH--NS observations in setting limits on the size of an extra spatial dimension.  Taking $\pm0.00003$~ms~y$^{-1}$ to be the nominal attainable precision for measurements of the rate of change of orbital period for any BH--NS system, then from eq.~(\ref{periodrate}), observations of a BH--NS system with those parameters could be used to set a 95\% confidence upper limit to the size of the extra dimension of $L<0.056\mu$m, about 800 times better than the current 95\% confidence upper limit of $L<44\mu$m from torsion-balance experiments, and probably considerably better than could be set by earth-based torsion-balance gravity experiments in the foreseeable future.

\section*{Acknowledgments}

We wish to thank Nemanja Kaloper for important discussions in the initial stage of this project.  We also thank Robert Canter, Patrick Huber, Jonathan Link, and Paul Wiita for insightful comments.  DM is supported in part by the U.S. Department of Energy, grant DE-FG05-92ER40677, task A.





\begin{thebibliography}{100}

\bibitem{Adelberger:2009zz}
Adelberger, E.~G. , Gundlach, J.~H. , Heckel, B.~R. , Hoedl, S. , and Schlamminger, S.\ 2009,
{Prog. Part. Nucl. Phys.}, 62, 102 

\bibitem{I1} 
Antoniadis, I., Arkani-Hamed, N., Dimopoulos, S., \& Dvali, G.\ 1998, Physics Letters B, 436, 257 

\bibitem{ADD}
Arkani-Hamed, N.,  Dimopoulos, S., \& Dvali, G.\ 1998, Physics Letters B, 429, 263

\bibitem{Belczynski2002}
Belczynski, K., Kalogera, V., \& Bulik, T.\ 2002, Astrophysical Journal, 572, 407

\bibitem{LHC}
Bleicher, M., and Nicolini, P.\ 2010, {J. Phys. Conf. Ser.}, 237, 012008

\bibitem{Emparan:2002px}
Emparan, R., Fabbri, A., and Kaloper, N.\ 2002, {JHEP}, 08, 043

\bibitem{Emparan:2002jp}
Emparan, R., Garcia-Bellido, J., and Kaloper, N.\ 2003, {JHEP}, 01, 079

\bibitem{Fitzpatrick2006}
Fitzpatrick, A.~L. , Randall, L, and Wiseman, T.\ 2006, {JHEP}, 11, 033

\bibitem{Gregory:2008rf}
Gregory, R.\ 2009, {Lect. Notes Phys.}, 769, 259

\bibitem{Hadjidemetriou1963}
Hadjidemetriou, J.~D.\ 1963, {Icarus}, 2, 440

\bibitem{Hadjidemetriou1966}
Hadjidemetriou, J.~D.\ 1966, {Icarus}, 5, 34

\bibitem{Johannsen2009}
Johannsen, T.\ 2009, {A\&A}, 507(2), 617

\bibitem{JohannsenPsaltisMcClintock2009}
Johannsen, T., Psaltis, D., \& McClintock, J.~E.\ 2009, Astrophysical Journal, 691, 997 

\bibitem{Kaluza}
Kaluza, T.\ 1921, {Sitzungsber. Preuss. Akad. Wiss. Berlin (Math. Phys. )}, 1921, 966, 1921

\bibitem{Kapner2007}
Kapner, D.~J., Cook, T.~S., Adelberger, E.~G., Gundlach, J.~H., Heckel, B.~R., Hoyle, C.~D., \& Swanson, H.~E.\ 2007, {Phys. Rev. Lett.}, 98, 021101

\bibitem{gressay} 
Kavic, M., Minic, D., \& Simonetti, J.~H.\ 2008, International Journal of Modern Physics D, 17, 2495 

\bibitem{Kavic:2008qb}
Kavic, M., Simonetti, J.~H., Cutchin, S.~E., Ellingson, S.~W., Patterson, C.~D.\ 2008,
JCAP, {0811}, 017

\bibitem{Klein}
Klein, O.\ 1926, {Z. Phys.}, 37, 895

\bibitem{Kudoh:2003xz}
Kudoh, H., Tanaka, T., and  Nakamura, T.\ 2003, {Phys. Rev.}, D68, 024035

\bibitem{Maldacena:1997re}
Maldacena, J.\ 1999, Trends in Theoretical Physics II, 484, 51

\bibitem{mathur}
Mathur, S.~D.\ 2009, {Class. Quant. Grav.}, 26, 224001

\bibitem{McWilliams:2009ym}
McWilliams, S.~T.\ 2010, {Phys. Rev. Lett.}, 104, 141601

\bibitem{Peters:1963ux}
Peters, P.~C., \& Mathews, J.\ 1963, {Phys. Rev.}, 131, 435

\bibitem{Peters:1964zz}
Peters, P.~C.\ 1964, {Phys. Rev.}, 136, B1224

\bibitem{Pfahl2005}
Pfahl, E., Podsiadlowski, P., \&  Rappaport, S.\ 2005, Astrophysical Journal, 628, 343

\bibitem{polch}
Polchinski, J.\ 1998, {String theory. Vol. 1: An introduction to the bosonic string},
Cambridge, UK: Univ. Pr.

\bibitem{PsaltisXTEJ1118}
Psaltis, D.\ 2007, {Phys. Rev. Lett.}, 98(18), 181101

\bibitem{RS2}
Randall, L., \& Sundrum, R.\ 1999, Physical Review Letters, 83, 4690 

\bibitem{Rattazzi2006}
Rattazzi, R. 2006, \newblock {Cargese Lectures on Extra Dimensions}

\bibitem{Sivardiere1985}
Sivardi\`ere, J.\ 1985, {Eur. J. Phys.}, 6, 245

\bibitem{Taylor:1982zz}
 Taylor, J.~H., \& Weisberg, J.~M.\ 1982, Astrophysical Journal, 253, 908 

\bibitem{Taylor:1989sw}
Taylor, J.~H., \& Weisberg, J.~M.\ 1989, Astrophysical Journal, 345, 434 

\bibitem{Weisberg:2004hi}
Weisberg, J.~M., \& Taylor, J.~H.\ 2005, Binary Radio Pulsars, 328, 25 

\bibitem{Yoshino:2008rx}
Yoshino, H.\ 2009, {JHEP}, 01, 068 

\end{thebibliography}
\end{document}